\documentclass[prl,twocolumn,superscriptaddress]{revtex4}
\usepackage{epsfig,amsmath,amssymb,graphicx,color,calc}

\newcommand{\beq}{\begin{equation}}
\newcommand{\eeq}{\end{equation}}
\newcommand{\ba}{\begin{align}}
\newcommand{\ea}{\end{align}}
\renewcommand{\phi}{\varphi}

\begin{document}

\title{A microscopic mean-field theory of the jamming transition}

\author{Hugo Jacquin}
\affiliation{Laboratoire Mati\`ere et Syst\`emes Complexes, UMR CNRS 7057,
Universit\'e Paris Diderot -- Paris 7, 10 rue Alice Domon et L\'eonie 
Duquet, 75205 Paris cedex 13, France}

\author{Ludovic Berthier}
\affiliation{Laboratoire Charles Coulomb, 
UMR 5221 CNRS and Universit\'e Montpellier 2, Montpellier, France}

\author{Francesco Zamponi}
\affiliation{Laboratoire de Physique Th\'eorique, 
\'Ecole Normale Sup\'erieure, UMR CNRS 8549, 24 Rue Lhomond, 75231 
Paris Cedex 05, France}

%\date{\today}

\begin{abstract}
Dense particle packings acquire rigidity through 
a nonequilibrium jamming transition commonly observed in materials
from emulsions to sandpiles. 
We describe athermal packings and their observed geometric phase transitions 
using equilibrium statistical mechanics and develop  
a fully microscopic, mean-field theory of the jamming transition
for soft repulsive spherical particles. We derive analytically 
some of the scaling laws and exponents characterizing the transition
and obtain new predictions for microscopic correlation functions 
of jammed states that are amenable to experimental verifications,
and whose accuracy we confirm using computer simulations.
\end{abstract}

\maketitle

About 50 years ago, Bernal~\cite{bernal} used 
dense disordered sphere packings as model systems to understand 
the liquid state, at a time where the statistical mechanics 
of liquids was still in its infancy. Today, the idea that
jammed materials share deep similarities with dense liquids
and glasses remains popular~\cite{liunagel}. 
However, while liquid state theory
grew as a cornerstone of theoretical physics~\cite{hansen}, 
no equivalent theory is available for jammed matter,
because this is a nonequilibrium, amorphous, athermal state
of matter---a theoretical
challenge overlooked by Bernal. Thus, despite
intense research activity~\cite{maxime,wyart,PZ} with a large body of 
numerical and experimental 
observations~\cite{vanhecke,ohernmodel}, it is not yet clear 
what is the appropriate theoretical framework to understand
dense athermal packings and the intriguing phase transitions
they undergo, although foams, pastes and emulsions are familiar materials.

We address the purely geometrical packing 
problem of soft spheres,
and suggest to study first their statistical mechanics 
at finite temperatures, $T$, before
taking the $T \to 0$ limit where jamming occurs.
A similar approach is frequently used in 
combinatorial optimization problems~\cite{optimization}, because 
powerful statistical mechanics tools can then be used 
in a context where they are not a priori relevant~\cite{KK}.
We investigate the statistical mechanics of the system at $T \geq 0$
using mean-field theory~\cite{MP,PZ}, and develop 
a fully microscopic theoretical scheme 
to predict the structure of non-equilibrium
configurations of soft repulsive spheres at zero~\cite{donev}
and finite~\cite{vestige} temperatures.

Our microscopic approach is thus markedly different from recent theoretical 
works~\cite{maxime,wyart}, which are based on phenomenological and
scaling considerations.
Similarly to Landau-Ginzburg theory of phase transitions, our aim is to 
derive, from first principles, 
the correct qualitative description of the transition, and
accurate quantitative predictions for several observables.
However, as a mean-field theory, our approach 
does not describe well all fluctuations near the transition,
and the associated scaling laws~\cite{ohernmodel,wyart}: 
these should be included by developing a renormalization group 
treatment around mean-field 
theory, which is currently under construction~\cite{RG}.

To make our approach concrete, 
we study an assembly of $N$ spherical particles of diameter $\sigma$
enclosed in a volume $V$ in three spatial dimensions, 
interacting with a soft repulsion
of finite range. To fix ideas we choose
\begin{equation}
V(r \leq  \sigma) = \epsilon (1-r/\sigma)^\alpha, \quad V(r > \sigma)=0,
\label{pot}
\end{equation}
with $r$ the interparticle distance, $\epsilon$ 
the strength of the repulsion, and $\alpha=2$ (harmonic repulsion).
Although several systems are described by a Hertzian
repulsion ($\alpha=\frac{3}{2}$), the harmonic model 
originally proposed to describe wet 
foams~\cite{durian} has become a paradigm in numerical studies 
of the $T=0$ jamming 
transition~\cite{ohernmodel,vanhecke}. It
was also studied at finite temperatures~\cite{tom,vestige}, 
and finds experimental realizations 
in emulsions and soft colloids.
The choice $\alpha=2$ is also technically more convenient, 
but we emphasize that our approach is easily generalized
to any repulsive potential. The model has two
control parameters: the temperature $T$,
and the fraction of the volume occupied by the 
particles in the absence of overlap: $\phi = \pi N \sigma^3/ (6V)$.
We set $\sigma$ and $\epsilon$ to unity.

Over the last decade, a number of numerical observations 
were reported for this model~\cite{vanhecke}.
A jamming transition is observed at $T=0$ at some critical
volume fraction, $\phi_j$, the   
density above which packings carry a finite density
of particle overlaps. Numerically, energy density, $e_{\rm gs}$, 
and pressure, $P$, are found
to increase continuously from zero above $\phi_j$
as power laws~\cite{ohernmodel}. The pair correlation function of
density fluctuations~\cite{hansen}, $g(r)$, 
develops singularities near $\phi_j$~\cite{donev},
which are smoothed by thermal excitations~\cite{vestige}. In particular,
$g( 1 ) = \infty$ at $\phi_j$ and $T=0$, which
implies that the density of contacts between particles, $z$, 
jumps discontinuously from $0$ to a finite value, $z_c$, 
at $\phi_j$.
Above $\phi_j$, $z$ increases algebraically with 
$\phi$~\cite{ohernmodel,vanhecke}. Thus, jamming
appears as a phase transition taking place in the absence of thermal 
motion, with a peculiar critical behaviour and observable
physical consequences~\cite{vanhecke}.

The success of our approach relies on its ability 
to accurately describe dense 
systems of harmonic spheres at very low $T$, which is
theoretically challenging~\cite{noi1}.
Simple liquid state theories, such as integral equations~\cite{hansen},
work well for dense systems only when $T$ is not too low~\cite{hugoSM}.
At lower $T$, numerical simulations~\cite{tom} indicate the appearance 
of a complex free energy landscape associated with slow dynamics, 
as found in glass-forming liquids.
Theoretically, the mode-coupling theory of glasses
can be applied~\cite{hugoPRE} but only gives limited 
dynamical insights, in particular failing to identify
the jamming transition~\cite{noi1}. 
For structure, a possible
path is a recently developed analytical approach 
based on replica calculations~\cite{monasson,MP}.
While generic in principle, the method requires in practice 
specific approximations. We have found that prior works
on Lennard-Jones~\cite{MP} and hard sphere~\cite{PZ} models
fail when directly applied to models of harmonic spheres 
near jamming~\cite{noi1}. 
The main technical 
achievement of the present paper is the derivation of a new analytical 
scheme to compute the structure and thermodynamics of the model
(\ref{pot}) over a range of parameters broad enough 
to allow the study of the $T=0$ jamming transition.

\begin{figure}
\psfig{file=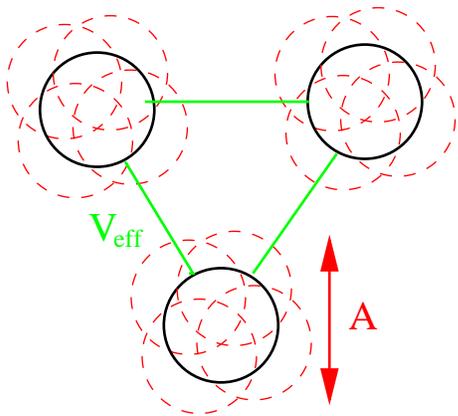,width=6.cm}
\caption{ 
Sketch of the derivation of the replicated 
free energy and effective potential in Eq.~(\ref{freem}). 
Each particle in the original liquid
is replicated $m$ times (dashed spheres). Assuming that the replicated
particles form a ``molecule'' of average cage size $A$, we trace 
out in the partition sum 
the degrees of freedom of $(m-1)$ copies of the liquid to 
obtain an effective one-component liquid (black spheres) interacting 
with an effective pair potential $V_{\rm eff}(r)$ 
(green lines).}\label{Veff} 
\end{figure}

We introduce $m$ copies of the system of harmonic spheres 
as a mathematical tool to probe its complex free energy 
landscape~\cite{monasson}, 
and develop approximations to study the statistical mechanics
of the replicated liquid using an effective potential 
approach~\cite{PZ}, as sketched in Fig.~\ref{Veff}. 
We make a Gaussian ansatz for the 
probability distribution of replicated particles
around the center of mass of 
the ``molecules'' shown in Fig.~\ref{Veff}, 
$\rho({\bf x}_1 \cdots {\bf x}_m) 
= \int d^3 {\bf X} \prod_{a=1}^m  (2\pi A)^{-3/2} 
e^{- ({\bf x}_a - {\bf X})^2 / (2A)}$, which defines the cage size $A$. 
Our central approximation is now performed, in which 
only two-body interactions between particles 
in copy $1$ induced by the coupling to the other $(m-1)$ copies
are retained, see Fig.~\ref{Veff}. 
Consider two ``molecules'', each composed of 
$m$ particles with positions 
$({\bf x}_1 \cdots {\bf x}_m)$ 
and $({\bf y}_1 \cdots {\bf y}_m)$: the effective potential
between the particles of replica 1
is obtained by averaging the total interaction 
$\sum_{a=1}^m V({\bf x}_a-{\bf y}_a)$
over the positions of particles within the $(m-1)$ remaining replicas:
\begin{equation}
\begin{split}
& e^{-\beta V_{\rm eff}({\bf x}_1-{\bf y}_1)}  \equiv \int d^3 {\bf x}_{2} 
d^3 {\bf y}_2 \cdots 
d^3 {\bf x}_m d^3 {\bf y}_{m} \\ & \times \bigg\{
 \rho( {\bf x}_1 \cdots {\bf x}_m) 
\rho({\bf y}_1 \cdots {\bf y}_m) 
\prod_{a=1}^m e^{-\beta V({\bf x}_a - {\bf y}_a)} \bigg\} \ .
\end{split}\end{equation}
Thanks to the Gaussian form of the integral the latter expression 
can be rewritten as follows:
\begin{equation}
\begin{split}
e^{-\beta V_{\rm eff}(r)} = \frac{ e^{-\beta V(r)} }{r \sqrt{4 \pi A}} 
\int_0^\infty \hskip-8pt du \left[ e^{-\frac{(r-u)^2}{4A}} - e^{-\frac{(r+u)^2}{4A}}
 \right] u \, q^{m-1}(u) \ ,
\end{split}
\label{met1}
\end{equation}
where $q(u)=  
\int d^3t \, e^{-\beta V(u-t)} \, e^{-t^2/(4A)}/(4 \pi A)^{3/2}$ 
has an explicit expression in terms of error functions, and 
$\beta =1/T$.
Finally, the free energy $F(m,A;\phi,T)$ is obtained by considering 
$V_{\rm eff}(r) - m V(r)$ as small, which becomes exact when $A \to 0$, 
see Eq.~(\ref{met1}), and doing standard
perturbation theory~\cite{hansen} 
around the liquid with potential $m V(r)$, 
which is equivalent to a liquid with potential
$V(r)$ at temperature $T/m$. 
We obtain an effective one-component system with a
free energy parametrized by $A$ and $m$:
\begin{equation}
\begin{split}
\label{freem}
&F( m,A;\varphi,T)   =  
F_{\rm harm}(m,A) + F_{\rm liq}(\varphi,\frac{T}{m}) \\
& -  
\frac{3 \phi T}{\pi} \int dr g_{\rm liq}(r, \phi, \frac{T}{m}) 
[ e^{-\beta [V_{\rm eff}(r) - m V(r)]} -1],
\end{split}
\end{equation}
where $F_{\rm liq}(\varphi,T)$ and $g_{\rm liq}(r,\phi,T)$ are
respectively the free energy and pair correlation function 
of the original (nonreplicated)
fluid, and $F_{\rm harm} = - \frac{3 T}{2} [ (m-1) \ln (2\pi A) + m - 1 + \ln m]$
is the ideal gas
contribution for the replicated system~\cite{MP}.
Thus the core of the approximation is 
embodied by the effective potential $V_{\rm eff}(r)$. Physically,
the presence of $(m-1)$ replicas induces 
near jamming a strong short-range {\it effective} 
attraction, similar in spirit to
depletion forces in colloid-polymer mixtures~\cite{PZ}.  

Our task becomes the study of a 
complicated effective fluid described by Eq.~(\ref{freem}). 
To simplify calculations we perform
a standard approximation, 
\begin{equation}
g_{\rm liq}(r,\phi,T) \equiv 
e^{-\beta V(r)} y (r, \phi, T) \approx  e^{-\beta V(r)} y (1,\phi, 0),
\end{equation}
well-suited to study the $T \to 0$ limit~\cite{hansen}.
To compute $F_{\rm liq}$ and $y(1,\phi,0)$ 
analytically we choose the hypernetted chain 
approximation~\cite{hansen}, although more elaborate 
closure relations~\cite{hansen} could be used. 
This could change slightly the location of the transition, but not 
its nature or the scaling predictions we derive.
Finally, to obtain concrete results for a given state point $(\phi, T)$, 
we minimize the free energy with respect to the 
cage size $A$ and, in the glass phase, 
to the replica number $m$~\cite{monasson}.

\begin{figure}
\psfig{file=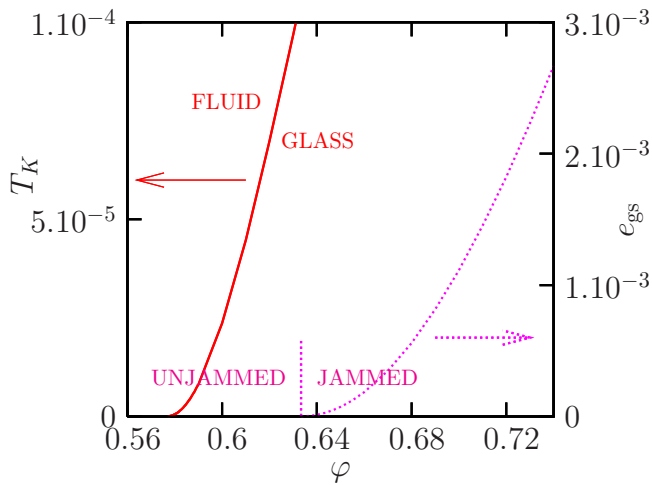,width=8.5cm}
\caption{
Theoretical phase diagram of soft repulsive spheres.
The glass transition temperature $T_K$ separates the 
liquid and glass phases with $T_K \sim (\phi - \phi_K)^2$ near 
$\phi_K \approx 0.577$.
At $T=0$, the glass jams under compression across 
$\phi_{\rm gcp} \approx 0.633$, above which 
no glass state with no particle overlap exists at $T=0$.
Thus, the ground state energy $e_{\rm gs}$ increases
continuously from 0 as $e_{\rm gs} \sim (\phi - \phi_{\rm gcp})^2$.
} \label{figphd}
\end{figure}
 
We first determine the location of the transition between the 
fluid and glass phases, signaled by the appearance of a 
free energy minimum with $m<1$~\cite{MP,monasson}, see Fig.~\ref{figphd}. 
A finite temperature glass transition, $T_K$, emerges continuously
from zero above  $\phi_K \approx 0.577$,
as $T_K \sim (\phi - \phi_K)^2$. We obtain 
the full thermodynamic behaviour in the glass phase (notably 
energy, pressure, specific heat, glass fragility) which compares
qualitatively well with numerical results~\cite{tom}. 
In particular, the ground state energy and pressure remain
zero across $\phi_K$, showing that just above $\phi_K$,
$T=0$ {\it glasses are not jammed}. In these glassy states, 
like in a hard sphere crystal, particles can vibrate near well-defined
(but random) positions, and the system is not jammed~\cite{PZ}.

We now concentrate on the $T \to 0$ limit at large volume 
fraction in the glass phase. We obtain the ground state energy 
shown in Fig.~\ref{figphd}. 
As found in simulations~\cite{ohernmodel}, it grows
continuously from zero above a critical packing fraction,
$e_{\rm gs} \sim (\phi - \phi_{\rm gcp})^2$, so
that the pressure increases linearly,
$P \sim (\phi - \phi_{\rm gcp})$. 
The `glass close packing'~\cite{PZ}, $\phi_{\rm gcp}$, represents 
in our calculations the largest 
density where $T=0$ glasses 
with no particle overlap exist. Within
the present approximation we obtain $\phi_{\rm gcp} = 0.633 > \phi_K$. 
Thus, from the sole knowledge of $V(r)$ in 
Eq.~(\ref{pot}), our theory predicts the existence 
and location of a jamming transition deep in the glass phase, 
and accounts for its critical nature.
Since a large number of metastable states exist
in the glass phase, our approach also
directly explains the strong protocol dependence
of the critical jamming density $\phi_j$ observed in 
simulations~\cite{ohernmodel,jline}, which get 
arrested in nonequilibrium amorphous states, and thus jam
at a critical packing fraction $\phi_j < \phi_{\rm gcp}$.
However, the results we obtain near $\phi_{\rm gcp}$ below are found to
hold for any metastable glass, and therefore also hold
near any protocol-dependent $\phi_j$.

We now turn to the calculation of the pair correlation function  
near $\phi_{\rm gcp}$. Within our approximation, $g(r)$ is 
directly related to the effective potential:
\begin{equation}
g(r) = e^{- \beta V_{\rm eff}(r)} y(1,\phi,0),
\end{equation}
and comes as a direct result of the free energy minimization.
We concentrate on the physics of interparticle contacts and thus
focus on distances $r \approx 1$, in the vicinity of the
jamming transition $(T \ll 1, \phi \approx \phi_{\rm gcp})$. 

\begin{figure}
\psfig{file=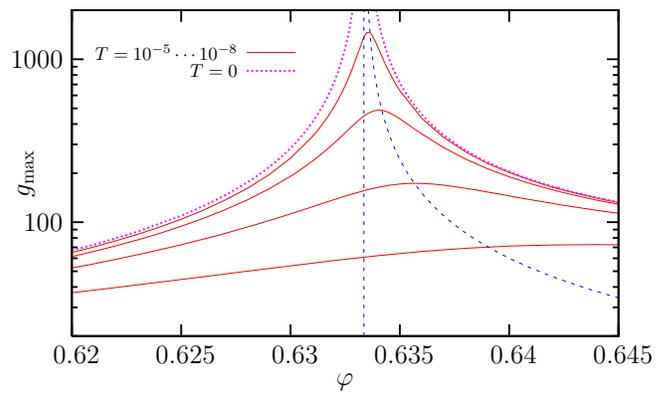,width=8.5cm,clip}
\caption{
Evolution of the maximum of the 
pair correlation function near contact with $T$ and $\phi$. 
While $g_{\rm max}$ diverges 
on both sides of the transition at $T=0$ as $g_{\rm max} 
\sim |\phi - \phi_{\rm gcp}|^{-1}$, 
this divergence becomes a smooth maximum at finite $T$ 
near the transition whose position shifts as $\sqrt{T}$
(dashed line), as observed numerically~\cite{donev,vestige}
and experimentally~\cite{vestige,tapioca}.
} \label{vestige}
\end{figure}

\begin{figure}
\psfig{file=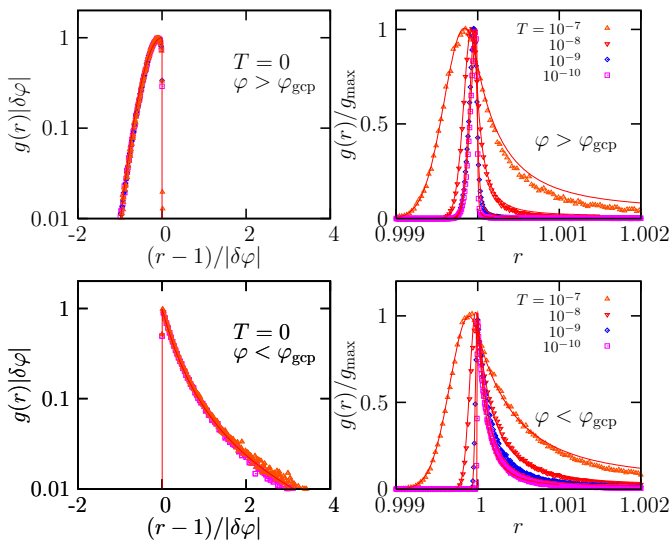,width=8.8cm,clip}
\caption{
The pair correlation near jamming
predicted by theory (full lines) and measured in numerical simulations
(symbols).  
Left panels: scaling behaviour at $T=0$ above (top) and below
(bottom) the jamming transition showing the convergence of the first
peak near $r=1$ to a delta function with asymmetric scaling functions on both 
sides of the transition. Right panels: 
the first peak of the pair correlation broadens when 
$T$ increases at constant $\phi$ above (top: $\delta \phi = 
2.8 \cdot 10^{-4}$) and below the transition
(bottom $\delta \phi = - 3.5 \cdot 10^{-4}$). 
To ease visualization, we show the
evolution of $g(r)/g_{\rm max}$, where $g_{\rm max}$ can 
be read from Fig.~\ref{vestige}.
} \label{marin}
\end{figure}

At $T=0$, we find that $g(r)$ develops a diverging peak near contact,
which obeys the following scaling law:
\begin{equation} 
g(r) \approx |\delta \phi|^{-1} {\cal F}_\pm \left[ \frac{r-1}{|\delta \phi|}
\right ], 
\label{T=0scaling}
\end{equation}
where ${\cal F}_\pm(x)$ are asymmetric scaling functions
which depend on the sign of $\delta \phi \equiv \phi - \phi_{\rm gcp}$ and
can be computed analytically.
In particular, $\log {\cal F}_+(x) \propto -x^2$ and 
${\cal F}_-(x) \propto x^{-2}$
when $x \gg 1$. Note that ${\cal F}_{-}(x)$ can be derived 
using the hard sphere potential~\cite{PZ}.
The scaling (\ref{T=0scaling}) means that the peak 
height, $g_{\rm max}$,
diverges as $|\delta \phi|^{-1}$
on both sides of the 
transition at $T=0$, see Fig.~\ref{vestige}, while its
width vanishes as $|\delta \phi|$. This behaviour was found 
in simulations~\cite{donev}.
In Fig.~\ref{marin} we show that numerical results
not only obey the scaling form in Eq.~(\ref{T=0scaling}), 
but also that the asymmetric shape of the scaling
functions compares extremely well with our theoretical
predictions.

At very low, but finite temperature, the peak divergence is smoothed 
by thermal fluctuations, which was the focus of a recent
study~\cite{vestige}.
In Fig.~\ref{vestige} we show the predicted smooth 
evolution of $g_{\rm max}$
when $\phi_{\rm gcp}$ is crossed at finite $T$. A nonmonotonic 
evolution with density is obtained, as in 
experiments~\cite{vestige,tapioca}. The position of the maximum evolves 
with $T$ with a scaling in perfect agreement
with numerical work~\cite{vestige}. We go further and 
predict the thermal broadening of $g(r)$, see 
Fig.~\ref{marin}. We obtain nearly perfect agreement
of theory with simulations over several 
decades of temperatures with no adjustable parameter.
The full scaling of $g(r)$ near contact, as a function
of $T$ and $\delta\phi$ near the jamming transition, 
is the main new achievement of the present work. 

Finally we obtain the number of contacts per particle by integration:
$z = 24 \phi \int_0^\infty dr r^2 g(r)$. The diverging peak described 
by Eq.~(\ref{T=0scaling}) gives a discontinuous jump of $z$ from 
0 to $z_c=6$, the celebrated isostatic value, 
at $\phi_{\rm gcp}$, as observed~\cite{bernal,ohernmodel,vanhecke}
and already derived in~\cite{PZ}.
Above the transition we find $z - z_c \propto (\phi - \phi_{\rm gcp})^\gamma$ 
with 
$\gamma=1$. The exponent is in quantitative disagreement with the
observed $\gamma=\frac{1}{2}$~\cite{ohernmodel}. 
Indeed, this exponent has been related to the presence
of fluctuations~\cite{wyart} that are presumably not well captured by
our mean-field theory, indicating that more detailed
calculations (possibly based on renormalization
group~\cite{RG}) should be developed 
to predict $g(r)$ over a broader range of interparticle distances.

Our results show that the fully nonequilibrium problem of 
soft particle packings, relevant to understanding
the mechanical properties of many soft materials, can be 
successfully address using 
equilibrium statistical mechanics tools. As an application
we have developed a many-body theory of the jamming
transition of soft repulsive spheres which satisfactorily  
derives, from first principles, 
the existence and location of a jamming transition, and some 
of its peculiar critical behaviour, and makes new predictions
for correlation functions of jammed states. Our approach is general
enough that it can be systematically improved and generalized to various
models, such that new, or more precise predictions could be made, 
hopefully fostering more numerical or experimental work. 
While Bernal saw packings as simplified models for atomic liquids, 
it is equally useful to consider packings as a special 
class of disordered ground states.

We thank P. Chaudhuri for providing the 
configurations used in numerical verifications of theoretical predictions 
in Fig.~\ref{marin}. 
H. Jacquin acknowledges financial support from Capital Fund 
Management (CFM) Foundation.
L. Berthier is partially funded by 
ANR Dynhet and R\'egion Languedoc-Roussillon.  

\bibliography{scibib}

\begin{thebibliography}{99}

\bibitem{bernal} J. D. Bernal, 
Nature {\bf 183}, 141 (1959);
J. D. Bernal and Mason, Nature {\bf 188}, 910 (1960).

\bibitem{liunagel} A. J. Liu and S. R. Nagel,
Nature {\bf 396}, 21 (1998).

\bibitem{hansen} J. P. Hansen and I. R. McDonald, \textit{Theory of 
Simple Liquids} (Elsevier, Amsterdam, 1986).

\bibitem{maxime} 
M. Clusel, E. I. Corwin, A. O. N. Siemens, and J. Brujic, 
Nature {\bf 460}, 611 (2009);
C. Song, P. Wang, and H. A. Makse, Nature {\bf 453}, 629 (2008).

\bibitem{wyart} M. Wyart, L. Silbert, S. R. Nagel, and 
T. Witten, Phys. Rev. E {\bf 72}, 051306 (2005). 

\bibitem{PZ} G. Parisi and F. Zamponi, 
Rev. Mod. Phys. {\bf 82} 789 (2010).

\bibitem{ohernmodel} C. S. O'Hern, S. A. Langer, 
A. J. Liu, and S. R. Nagel, Phys. Rev. Lett. {\bf 88}, 075507 (2002).

\bibitem{vanhecke} M. van Hecke, 
J. Phys. Condens. Matter {\bf 22}, 033101 (2010);
A. J. Liu, S. R. Nagel, W. van Saarloos, and M. Wyart, {\tt arXiv:1006.2365}.

\bibitem{optimization}
M. M\'ezard and A. Montanari, {\it Information, Physics, and Computation}
(Oxford University Press, Oxford, 2009).

\bibitem{KK} F. Krzakala and J. Kurchan,
Phys. Rev. E {\bf 76}, 021122 (2007).

\bibitem{MP} M. M\'ezard and G. Parisi,
J. Chem. Phys. {\bf 111}, 1076 (1999).

\bibitem{donev}
A. Donev, S. Torquato, and F. H. Stillinger,
Phys. Rev. E {\bf 71}, 011105 (2005);
L. Silbert, A. J. Liu, and S. R. Nagel,
Phys. Rev. E {\bf 73}, 041304 (2006). 

\bibitem{vestige} Z. Zhang {\it et al.}, 
%N. Xu, D. T. N. Chen, P. Yunker, 
%A. M. Alsayed, K. B. Aptowicz, P. Habdas, A. J. Liu, 
%S. R. Nagel, and A. G. Yodh, 
Nature {\bf 459}, 230 (2009). 

\bibitem{RG} 
M. Castellana {\it et al.}, Phys. Rev. Lett. {\bf 104}, 127206 (2010);
C.~Cammarota, G.~Biroli, M.~Tarzia, G.~Tarjus, Phys. Rev. Lett. {\bf 106}, 115705 (2011).

\bibitem{durian} D. J. Durian,
Phys. Rev. Lett. {\bf 75}, 4780 (1995).

\bibitem{tom} L. Berthier and T. A. Witten, 
EPL {\bf 86},100001 (2009); Phys. Rev. E {\bf 80}, 021502 (2009).

\bibitem{noi1}
L.~Berthier, H.~Jacquin and F.~Zamponi, J.Stat.Mech. (2011) P01004.

\bibitem{hugoSM} 
H. Jacquin and L. Berthier, Soft Matter {\bf 6}, 2970 (2010). 

\bibitem{hugoPRE} L. Berthier, E. Flenner, H. Jacquin, and G. Szamel,
Phys.~Rev.~E {\bf 81}, 031505 (2010);
W. T. Kranz, M.~Sperl, and A.~Zippelius,
Phys.~Rev.~Lett. {\bf 104}, 225701 (2010).

\bibitem{monasson} R. Monasson, Phys. Rev. Lett. {\bf 75}, 2847 (1995).

\bibitem{jline}
P. Chaudhuri, L. Berthier, and S. Sastry,
Phys. Rev. Lett. {\bf 104}, 165701 (2010).
 
\bibitem{tapioca}
X. Cheng, Phys. Rev. E {\bf 81}, 031301 (2010).

\end{thebibliography}

\bibliographystyle{Science}

\end{document}